\documentclass[a4paper,fleqn,usenatbib]{mnras}
\usepackage{amssymb,amsmath,graphicx,psfrag,mathtools}
\usepackage[T1]{fontenc} 
\usepackage{ae,aecompl} 
\usepackage{url}
\usepackage{revsymb}
\hypersetup{draft} 
\title[FRB in Globular Clusters]{A FRB in a Globular Cluster: Why Is This
Neutron Star Different From (Almost) All Other Neutron Stars?}
\author[J. I. Katz]{
J. I. Katz,$^{1}$\thanks{E-mail katz@wuphys.wustl.edu} 
\\
$^{1}$Department of Physics and McDonnell Center for the Space Sciences,
Washington University, St. Louis, Mo. 63130 USA 
}
\date{Accepted XXX.  Received YYY; in original form ZZZ} 
\pubyear{2020} 
\date{\today}
\begin{document} 
\label{firstpage} 
\pagerange{\pageref{firstpage}--\pageref{lastpage}} 
\maketitle 
\begin{abstract}
	Most Fast Radio Burst (FRB) models are built from comparatively
	common astronomical objects: neutron stars, black holes and
	supernova remnants.  Yet FRB sources are rare, and most of these
	objects, found in the Galaxy, do not make FRB.  Special and rare
	circumstances may be required for these common objects to be sources
	of FRB.  The recent discovery of a repeating FRB in a globular
	cluster belonging to the galaxy M81 suggests a model involving a
	neutron star and a close binary companion, likely a white dwarf;
	both neutron stars and close binaries are superabundant in globular
	clusters.  Magnetic interaction is a plausible, though unproven,
	mechanism of acceleration of relativistic particles that may
	radiate coherently as FRB.  In such a model the energy source is the
	orbital kinetic energy, and not limited by the magnetostatic energy
	of a neutron star.  Double neutron star binaries cannot be the
	observed long-lived repeating FRB sources, but might make much
	shorter lived sources, and perhaps non-repeating FRB.
\end{abstract}
\begin{keywords} 
radio continuum, transients: fast radio bursts, stars: neutron, stars: binary
\end{keywords} 
\section{Introduction}
The localization \citep{Ki21} of FRB 20200120E to a globular cluster in the
galaxy M81 is an important clue to the origin and mechanism of FRB.  It has
long been known \citep{C75,K75} that Galactic globular clusters are
superabundant in neutron star binary X-ray sources.  They also contain many
``recycled'' pulsars with decayed magnetic fields, spun up to millisecond
periods by accretion from close binary companions \citep{S10}.  Some of
these companions have survived in bound orbits but others appear to have
dissipated.


Neutron stars are natural components of any model of FRB because of their
large magnetic fields, high-brightness coherent emission as pulsars, and
compact size, and soft gamma repeaters (SGR) have been suggested as their
sources by many \citep{K18}.  But there are a large number of neutron stars
and dozens of SGR in the Galaxy, and to date only one has been the source of
a FRB (or a FRB-like event, FRB 200428, orders of magnitude less energetic
than ``cosmological'' FRB): SGR 1935$+$2154
\citep{B20b,C20b,Li20,M20,Ri20,T20}.

This is evidence that ``ordinary'' neutron stars are not the principal
source of FRB \citep{K20a}; if neutron stars make
FRB, those that do, either directly as super-giant pulsar pulses, indirectly
as stimulants of surrounding supernova remnants or in some other manner,
must be distinguished in some way from other neutron stars.  The distinction
must be based on more than magnetic field or rotation rate, because the
Galaxy contains many neutron stars that don't make FRB but fill those
parameter spaces.  Age might be
relevant; FRB 121102 may be surrounded by a very young, dense, rapidly
changing and strongly magnetized supernova remnant (SNR) with varying
dispersion measure (DM) and rotation measure (RM) \citep{K21a}, but other
FRB do not appear to be embedded in dense, strongly magnetized plasma, and
show no evidence of a near-source contribution to either DM or RM.

How may the globular cluster environment naturally produce a rare subclass
of neutron stars that plausibly, or at least possibly, make FRB in a manner
that other neutron stars do not?  How does this subclass differ from the
more numerous globular cluster neutron stars that do {\it not\/} make FRB?
Here I outline some fragmentary ideas that address these questions. 
They are based on the hypothesis that an interacting binary is an environment
favorable for making FRB.  As with other FRB models \citep{P19}, I attempt
only to show the {\it possibility\/} of emitting the observed FRB; the
failure to explain radio emission of pulsars 54 years after their discovery
shows the difficulty of understanding coherent emission processes.
\section{How Globular Clusters are Special}
Globular clusters are superabundant in neutron stars, and in close neutron
star binaries (some of which are no longer binaries, as the nondegenerate
companions have been dissipated by accretion and excretion).  This is the
result of the clusters' high stellar densities \citep{P03}.

It is necessary to find circumstances that distinguish a few, rare, neutron
stars from the many other neutron stars in globular clusters that are low
mass X-ray binaries, recycled pulsars or entirely inactive.  Presence in an 
asynchronous binary system implies interaction, in analogy to the Jupiter-Io
interaction, and at least the possibility of particle acceleration and
coherent emission \citep{MZ14,MZV20}.  The requirement that the model occur
rarely, while consisting of common ingredients, points to a short lifetime
in its active (FRB-emitting) phase.

These conditions may be satisfied by a model in which a young neutron star
with a large magnetic moment $\mu \sim 10^{33}$ Gauss-cm$^3$ interacts
magnetically with a close binary companion.  The low magnetic moments of
recycled pulsars in globular clusters are attributed to field decay at their
presumptive great ages, and do not exclude the possibility of birth with
fields $B \sim 10^{15}\,$G as large as those of SGR.  {In considering
exceptional objects low, but non-zero, ongoing rates of neutron star
formation by merger of double white dwarf binaries may be considered.}

A feature of a FRB model involving a {\it binary\/} neutron star is that
the orbital kinetic energy, equal to the orbital gravitational binding
energy, may be available to power FRB emission.  This may satisfy the
energy requirements of $10^{47}$--$10^{49}\,$ergs \citep{Ma20} that exceed
the magnetostatic energies $\sim 10^{47}\,$ergs of even the most strongly
magnetized known neutron stars, even apart from the unknown efficiency of
tapping magnetostatic energy.
\section{Parameters}
The lifetime of a close binary is limited by its emission of gravitational
radiation, which, assuming constant masses (no mass transfer, the companion
not overflowing its Roche lobe) is
\begin{equation}
	\label{tGR}
	t_{GR} = {5 \over 256}{c^5 \over G^3}{R^4 \over M_1 M_2 (M_1+M_2)},
\end{equation}
where $R$ is the separation of the stars and $M_1$ and $M_2$ their masses.
Only a degenerate companion is compact enough to permit a small $R$ and a
short $t_{GR}$ {in the binary's active phase as a FRB source}.  The
evolution of close neutron star-white dwarf binaries has recently been
discussed by \citet{CTHC21}.

{In order that a neutron star-white dwarf binary evolve to separations
small enough that interaction is possible, it must be formed with $R$ small
enough that $t_{GR}$ is less than the age of the Galaxy; Eq.~\ref{tGR}
implies that this requires the initial separation $R \lesssim 2 \times
10^{11}\,$cm.  A tighter constraint is imposed by the requirement that
$t_{GR}$ be less than the decay time of the neutron star's magnetic field;
this time scale is uncertain, but \citet{V13} indicate values of
$10^5$--$10^6\,$y, which would require $R \lesssim 2 \times 10^{10}\,$cm.
Three-body encounters among stars with speeds ${\cal O}(v_{rms})^2$, where
$v_{rms} \approx 20\,$km/s is a globular cluster's escape speed, produce
binaries with $R \sim G M_\odot/v_{rms}^2 \sim 3 \times 10^{13}\,$cm.
Suitable binaries can only be produced by tidal captures or collisions, in
which a neutron star passes close to (or enters the envelope of) a main
sequence star that later evolves to a white dwarf.}

The need to keep $R$ small and minimize $t_{GR}$ to make the systems rare,
while requiring that $t_{GR}$ not be less than observed repeating FRB source
lifetimes, pins $R$ {when the binary is an active FRB emitter} in the
range
\begin{equation}
	\label{R}
	1.3 \times 10^9\,\text{cm} < R \lesssim 5 \times 10^9\,\text{cm},
\end{equation}
where we have adopted $M_1 = 1.4M_\odot$ for the neutron star and $M_2 = 0.7
M_\odot$ for a white dwarf companion.  The steep dependence of $t_{GR}$ on
$R$ (Eq.~\ref{tGR}) and of the interaction energy $\cal E$ on $R$
(Eq.~\ref{E}) reduce the uncertainty of the upper bound, while $t_{GR} > 8$ y
(FRB 121102) gives the firm lower bound (weakly dependent on the assumed
masses).

The gravitational radiation lifetime is related to the orbital period
\begin{equation}
	P_{orb} = 2\pi \sqrt{R^3 \over G(M_1+M_2)}
\end{equation}
by
\begin{equation}
	P_{orb} = 2 \pi t_{GR}^{3/8} \left({256 G^{5/3} M_1 M_2 \over
	5 c^5 (M_1+M_2)^{1/3}}\right)^{3/8} = 18.5\left({t_{GR} \over
	\text{10 y}}\right)^{3/8}\ \text{s}.
\end{equation}
The orbital separation $R$ may be restated in terms of $t_{GR}$
\begin{equation}
	\label{RGR}
	R = 1.34 \times 10^9 \left({t_{GR} \over \text{10 y}}\right)^{1/4}\
	\text{cm}.
\end{equation}
For the assumed mass ratio, the radius of the Roche lobe of the less massive
companion $R_{WD} \approx 0.3R$.  Taking $R_{WD} = 7 \times 10^8$ cm
increases the lower bound to $R > 2.3 \times 10^9$ cm and $P_{orb} > 40\,$s.
If the white dwarf is semi-detached these lower bounds are approximate
values of the parameters.  It may be only a coincidence, but this $P_{orb}$
is close to the peak (60 s) of the distribution of burst intervals during an
active phase of FRB 121102 \citep{K19}.

For these parameters the gravitational radiation lifetime of the orbit is,
very approximately,
\begin{equation}
	t_{GR} \sim 100\,\text{y}.
\end{equation}
$t_{GR}$ is much longer for less massive white dwarfs because $t_{GR}
\propto R^4/M_2$; their larger $R_{WD}$ and smaller $R_{WD}/R$ would imply
greater $R$.  The model does not predict a measurably short life expectancy.
Perhaps FRB activity gradually declines on the time scale $t_{GR}$ as the
white dwarf loses mass and $R$ increases.  A rough scaling, using the
mass-radius relation $R_{WD} \propto M_2^{-1/3}$ for low mass white dwarfs
and $R \sim R_{Roche}(M_1/M_2)^{1/3} \sim R_{WD}(M_1/M_2)^{1/3} \propto
M_2^{-2/3}$ is $t_{GR} \propto R^{11/2} \propto M_2^{-11/3}$.
\section{Interaction Energy}
\label{IE}
{The rarity of FRB sources permits, even compels, the consideration of
unusual objects.  Here I evaluate the interaction energy of a strongly
magnetized neutron star with a white dwarf companion.  Known neutron stars
in globular clusters have very small fields, their primordial fields having
decayed.  The hypothesis is that FRB are produced by very young neutron
stars whose fields have not yet decayed.  Such neutron stars must
constitute a tiny fraction of neutron stars in globular clusters, but that
is consistent with the rarity of FRB sources.}
\subsection{Slowly rotating neutron star, non-magnetized white dwarf}
The energy density of a neutron star's magnetic field at its companion is
$\sim \mu^2/(8 \pi R^6)$, where $\mu$ is the neutron star's magnetic moment.
The magnetostatic energy associated with interaction with an unmagnetized
companion of radius $R_c$ is
\begin{equation}
	\label{E}
	{\cal E} \sim {B^2 \over 8\pi}{4 \pi \over 3}R_c^3 \sim
	{\mu^2 R_c^3 \over 6 R^6} \sim 4 \times 10^{35} \mu_{33}^2\ 
	\text{ergs},
\end{equation}
where the numerical value assumes a Roche lobe filling companion with $R_c =
7 \times 10^8\,\text{cm} = 0.3R$, appropriate to the assumed parameters of a
NS-WD binary system, and $\mu_{33} \equiv \mu/10^{33}\,$G-cm$^3$.
\subsection{Slowly rotating neutron star, magnetized white dwarf}
{If the white dwarf is strongly magnetized, with moment $\mu_{WD}$, then
Eq.~\ref{E} is replaced by
\begin{equation}
	\label{EWD}
	{\cal E} \sim {\mu\mu_{WD} \over R^3} \sim 10^{38} \mu_{33}
	\mu_{WD33}\,\text{ergs},
\end{equation}
where $\mu_{WD33} = \mu_{WD}/10^{33}\,$G-cm$^3$.  While the largest known
neutron star $\mu_{33} \sim 1$, white dwarfs are known with $\mu_{WD33}$ as
large as $\sim 100$ \citep{FMG15}, permitting ${\cal E} \sim 10^{40}\,$ergs.
The pairing of a strongly magnetic neutron star with a strongly magnetic
white dwarf is likely to be rare, but so are FRB sources.
\subsection{Rapidly rotating neutron star}
If the neutron star has an angular rotation rate $\omega > 2 \pi c/R$ and
spin period $P_{spin} = 2 \pi/\omega$ then the companion is in its radiation
zone, the interaction is not magnetostatic, and Eq.~\ref{E} is replaced by
\begin{equation}
	\label{Erad1}
	{\cal E} \sim {2 \over 9}{\mu^2 R_c^3 \over R^6}\left({R\omega \over
	c}\right)^4.
\end{equation}
The requirement that the spindown age be $> 8\,$y (requiring that the model
describe FRB 121102, and perhaps all repeating FRB) then implies $P_{spin}
> 0.5 \mu_{33}\,$s and
\begin{equation}
	\label{Erad2}
	{\cal E} \sim {2 \over 9}{\mu^2 R_c^3 \over R^2 c^4}\left({
		2 \pi \over P_{spin}}\right)^4 \lesssim 5 \times 10^{35}
		\mu_{33}^{-2}\ \text{ergs}.
\end{equation}
This energy can be much greater than that of Eq.~\ref{E} if $\mu_{33} \ll
1$.  Its maximum value is given by Eqs.~\ref{Erad1}, \ref{Erad2} if $\omega
\sim 10^4/$s, the fastest a neutron star can spin, and $\mu_{33} \sim
10^{-3}$, the greatest $\mu_{33}$ permitted by an eight year lifetime for
such a fast-spinning neutron star, and is ${\cal E} \sim 5 \times 10^{41}
\,$ergs.  As for strongly magnetized but slowly rotating neutron stars
interacting with strongly magnetized white dwarfs (Eq.~\ref{EWD}), it posits
an unprecedented object, a very young millisecond neutron star with a
typical pulsar field $B \sim 10^{12}\,$G.  This is consistent with the
rarity of FRB sources.
\subsection{Comparison to known Galactic objects}
The proposed models have much larger $\mu$ and smaller $R$ than known
Galactic binary millisecond pulsars and close double white dwarfs.  As a
result, $\cal E$ (Eqs.~\ref{E}, \ref{EWD}, \ref{Erad1}) is greater by many
orders of magnitude, explaining why the models predict that these binaries
do not radiate observable FRB, even though they are much closer than M81.
\subsection{Dissipation}
There is no first-principles theory of how this energy may be dissipated and
radiated.  The neutron star's magnetic field interacts with the companion
star, perhaps drawn into the companion's surface layers if they are
convecting, or into a mass-transfer flow if the companion fills its Roche
lobe.  Dissipation by magnetic reconnection may require that the neutron
star's rotation be asynchronous with the orbital period, as are known
neutron star binaries.

In SGR the entire magnetosphere appears to relax during giant outbursts
\citep{K82}, and the energy density is so high that the released energy must
thermalize into a near-equilibrium pair-photon fireball with a thermal
spectrum \citep{K96}.  At the lower energy densities of FRB coherent
radiation by relativistic particles may occur, in analogy with pulsars.
However, the extreme ratios of FRB energies to those of pulsar giant pulses
indicate that FRB cannot be explained as an extrapolation of observed pulsar
giant pulses \citep{BC19}.  They are outliers, indicating a qualitatively
different origin \citep{K21b}.}
\section{Numbers of Sources}
{I first estimate the expected number of FRB sources in the assumed
model at redshifts $z \lesssim 1$, even though the FRB discovered in a
globular cluster \citep{Ki21} is in M81 at a distance of 3.6 Mpc, about 1000
times closer than $z = 1$.  The cross-section for one star to approach
another within a distance $R$ (at which dissipation may bind it
{\bf gravitationally}) is
\begin{equation}
	\sigma \approx \pi {GMR \over v_\infty^2} \sim 2 \times 10^{27}\
	\text{cm}^2,
\end{equation}
where $M$ is the sum of the masses (taken to be $2.1 M_\odot$), $v_\infty
\approx 20\,$km/s is a typical globular cluster virial or escape velocity,
{\bf I have taken $R \sim 10^{13}\,$cm for capture into a common envelope
by a red giant that will evolve into a white dwarf}, and strong
gravitational focusing ($v_\infty \ll \sqrt{GM/R}$) has been assumed.  {\bf
Direct captures by white dwarfs are less frequent because of their small
radii.}

If there are $N$ neutron stars in a cluster, all in its densest central
regions (as expected from gravitational relaxation) where
stellar density is $n$, $N_G$ major (mass $\sim M^*$) galaxies with $z
\lesssim 1$, $N_{GC}$ such globular clusters per galaxy, and the captured
tightly bound neutron stars are active FRB sources for a time $t_{FRB}$,
then the total number of active FRB sources in the Universe is
\begin{equation}
	\label{NFRB}
	N_{FRB} = \sigma v_\infty N_G N n N_{GC} t_{FRB} \sim 10^4,
\end{equation}
where we have taken $N_G = 10^9$ \citep{C16}, $N = 10^3$, {\bf the red giant
density} $n = 10^{-53}\,$cm$^{-3}$ (300/pc$^3$), $N_{GC} = 100$ and $t_{FRB}
= 100\,$y.  This number is uncomfortably low, but $t_{FRB}$ could be one or
more orders of magnitude greater; estimated decay times of neutron star
magnetic fields \citep{V13} are three or four orders of magnitude greater
than the $t_{FRB}$ assumed here, with a proportional increase in $N_{FRB}$.

The price paid for assuming a larger $t_{FRB}$ is a larger $R$ and smaller
$\cal E$.  This problem is mitigated if FRB are collimated into a solid
angle $\Omega$, reducing the energy per burst by the factor $\Omega/4\pi$.
The burst rate must be correspondingly increased, but it is not strongly
constrained (it may be limited by the time required to restore a tangled
metastable field configuration, plausibly the neutron star's rotation
period).  As in most FRB models, the instantaneous power, not the mean
power, is the energetic constraint; collimation can reduce the inferred
instantaneous power by a large factor, but does not affect the inferred
mean power if the beam is isotropically distributed on the sky.

FRB 20200120E poses a special problem because in Eq.~\ref{NFRB} $N_G$ must
be replaced by the number of galaxies within about $3.6 \times f^{1/3}$ Mpc,
the distance within which $1/f$ of the galaxies are as close to us as M81,
or closer, so that the hypothesis that M81 is a random accident cannot be
rejected at a confidence level $> 1-1/f$.  Taking $f = 10$, and ignoring
the selection effect that closer FRB are more likely to be observed, $N_G$
is replaced by $\sim 300$ (the density of galaxies in the Local Group is
much greater than their mean cosmological density) and $N_{FRB\,<8\text{Mpc}}
\sim 10^{-4}$, apparently inconsistent with the observation of FRB
20200120E.  This may be resolved because the bursts of FRB 20200120E had
fluxes and fluences comparable to those of typical cosmological FRB, yet its
proximity implies that they were about $10^6$ times less energetic.  This
permits larger values of $R$ and $t_{FRB}$ longer by several orders of
magnitude (just how many depends on which of the models of Sec.~\ref{IE} is
adopted).  The processes that create possible neutron star FRB sources in
globular clusters have recently been discussed by \citet{KPL21,LBK21}.}
\section{Distribution of Burst Intervals}
A source that emits in stochastically wandering directions, but with a
rotationally modulated anisotropic likelihood (mean beam pattern), displays
a wide distribution of burst intervals that peaks at its rotation period
although the bursts are not strictly periodic \citep{K19}.  The absence of
strict periodicity \citep{Z18,Li21} in FRB 121102 is evidence in favor of
rotational modulation of stochastically directed emission.  It cannot (by
itself) distinguish between stochastic wandering of the direction of beamed
emission \citep{K17} and temporally stochastic emission of isotropic
radiation.  However, isotropic emission models place much more severe
demands on the {\it instantaneous\/} radiated power and energetics, which is
an argument in favor of wandering beams.
\section{Periodic Modulation?}
Two repeating FRB sources (FRB 20180916B \citep{CHIME20a,PM20,Pi20,Pl20}
and FRB 121102 \citep{R20,C21}) show activity modulated with periods of
16.35 d and $\approx 160$ d, respectively.  It is not known if this behavior
is universal among repeating FRB or is particular to some fraction of them;
these two sources are well-observed, and therefore selection favored
detection of such periods in them over other, less well observed, repeating
FRB, even if the behavior is universal.

Several models of this behavior may be considered in the context of the
present model of FRB:
\begin{enumerate}
	\item Precession of the orbital plane of the neutron star-white
		dwarf binary, driven by a third, more distant, companion.
		If all three masses are comparable, then the period of the
		outer orbit is roughly the geometric mean of the orbital
		period of the close binary and the precession period:
		\begin{equation}
			P_{\text{3rd body}} \sim \sqrt{P_{orb} P_{precess}}
			\sim \text{1--10\,d}.
		\end{equation}
		There is no evident means of detecting the third body
		directly.  If $P_{orb}$ or the neutron star's spin period
		were detected, the presence of the third body might produce
		a measureable Doppler shift.
	\item If the neutron star's rotation is synchronous with its
		orbital period, its orientation may oscillate in phase with
		a frequency
		\begin{equation}
			\omega_{osc} \sim \sqrt{{\cal E}/I},
		\end{equation}
		where $I$ is the neutron star's moment of inertia
		\citep{JKR79}.  For $\cal E$ given by Eq.~\ref{E}
		$\omega_{osc} \sim 10^{-5} \mu_{33}$/s, roughly consistent
		with the observed periods of SGR 20180916B and SGR 121102.
		However, for the much larger values of $\cal E$ given by
		Eqs.~\ref{EWD} and \ref{Erad2}, that more readily explain
		FRB energetics, $\omega_{osc}$ is much too large to be
		consistent with the observed periods.
	\item If the neutron star is rotating at a frequency $\omega_r$
		then its rotation axis will precess as a result of the
		torque exerted by the companion on its rotational equatorial
		bulge.  The precession frequency (if $M_{WD} \sim M_{NS}$)
		\begin{equation}
			\omega_{pre} \sim {M_{WD}r_{NS}^5 \over I R^3}
			\omega_r \sim {M_{WD} \over M_{NS}}
			\left({r_{NS} \over R}\right)^3 \omega_r
			\lesssim 10^{-10} \omega_r,
		\end{equation}
		which is too slow to explain the observed periodicities for
		plausible values of the parameters.
\end{enumerate}
\section{Future Behavior}
If the binary is detached, then as gravitational radiation brings the stars
together their interaction will become stronger and FRB activity may
gradually increase on a time scale $\sim t_{GR}$.  This inference is
uncertain because it ignores any effect of possible changes in the neutron
star's rotation period.  Dissipation may synchronize the neutron star's
rotation to the orbital period, as happens to white dwarfs in polars
\citep{JKR79}, ending dissipation and likely terminating FRB activity.

The evolution of semi-detached binaries, with the white dwarf losing mass to
the neutron star, is discussed by \citet{vH11,B17,Z20,B21,Y21}.  If the
white dwarf is much less massive, mass transfer may occur on the slow time
scale $t_{GR}$, as in recycled and ``black widow'' pulsars.  As the white
dwarf loses mass $R_c \approx R_{WD} \propto M_2^{-1/3}$ and $R \propto
M_2^{-2/3}$, so the interaction energy will decay on a time scale
$\sim t_{GR}$.  If the neutron star's rotation is not already synchronous,
synchronization may accelerate the decay of activity.  However, many such
binaries are known in the Galaxy, and they do not emit FRB.  Neutron star
binaries with more massive white dwarf companions are not known, and may
have very short lifetimes terminated by dynamical mass transfer, a burst of
gravitational waves, and a possible SN-like event or collapse to a black
hole.  Such a process would last much longer (seconds, and perhaps orders of
magnitude longer) than the ms duration of non-repeating FRB, but might
explain non-repetition.
\section{Discussion}
Explanation of repeating FRB in this model must reconcile the energies of
Eqs.~\ref{E}, \ref{EWD}, \ref{Erad2} with FRB energetics.  \citet{G18} report
bursts of FRB 121102 with fluences up to $\sim 1$ Jy-ms.  At its redshift of
0.193 this corresponds to an isotropic-equivalent energy $E_{iso} \sim
10^{39}\,$ergs, much greater than the interaction energy of Eq.~\ref{E}, and
uncomfortably close to the much greater interaction energies of
Eqs.~\ref{EWD}, \ref{Erad2}.  Even if a FRB is associated with a global
relaxation of the magnetic field in the interaction region of size $\sim
R_c$, the observed FRB fluences are possible only if their radiation is
narrowly collimated.  The arguments for collimation of FRB emission
\citep{K18,K19,K20b} occur in all models of repeating FRB.

In the simplest possible model of beamed radiation the FRB source emits
continuously, but its beam wanders on the sky \citep{K17}.  If it is
observed to have a duty factor $D$ by an unfavored observer (excluding the
possibility that the beam is preferentially directed towards the observer),
its solid angle of emission $\Omega \sim 4 \pi D\,$sterad.  For FRB 121102
$D \sim 2 \times 10^{-6}$ \citep{K19}, $\Omega \sim 2 \times 10^{-5}
\,$sterad, implying a Lorentz factor of the emitting charge bunches $\gamma
\gtrsim 200$, consistent with other constraints \citep{K20b} and with
relativistic kinematics with the assumed Lorentz factor \citep{K19}. 

{The required radiated burst energy $E \sim (\Omega/4\pi)E_{iso} \sim D
E_{iso} \sim 10^{33}\,$ergs.  This is consistent with the interaction
energies Eqs.~\ref{E}, \ref{EWD}, \ref{Erad2}, provided the magnetic field
in the interaction region can relax on the ms time scale of the FRB and
radiate coherently into a collimated beam, with its duration shortened from
$R_c/c \sim 25\,$ms to the observed $\sim 1\,$ms.  The slowly rotating model
with an unmagnetized companion requires an efficiency of coherent emission
of $\sim 1\%$ (Eq.~\ref{E}), but the magnetized companion and fast rotating
models only require efficiencies $\sim 10^{-9}$--$10^{-7}$ for their most
favorable parameters.  Pulsars turn spindown energy into coherent emission
with greater efficiency.}

If the companion were a neutron star \citep{T13,W16,DE17}, $\cal E$ would be
several orders of magnitude less than for a WD companion unless $R$ were
$\sim$ 10--100 times smaller than assumed here (Eq.~\ref{R}).  Such a small
$R$ would imply $t_{GR}$ $\sim$ $10^4$--$10^8$ times less, $\ll 1\,$y,
inconsistent with the observed lifetimes of repeating FRB.  It would be
consistent with non-repeating FRB or yet-unobserved very short-lived
(minutes--days) repeating FRB.  The hypothesis that non-repeaters are close
NS-NS binaries while repeaters are more distant NS-WD binaries might explain
the bimodal distribution of the duty factor $D$ \citep{K17,K18,K19}, with
upper bounds for apparently non-repeating FRB several orders of magnitude
less than the values observed for repeaters.  Then non-repeaters would be
associated with merging neutron stars, kilonov\ae\ and gravitational wave
events.  Testing that association would require large solid-angle or all-sky
sensitivity to FRB, such as provided by STARE2 \citep{B20a} or the proposed
lunar FRB scattering observatory \citep{K20c}, unfortunately at less
sensitivity than a high-gain focussing telescope.  The likelihood that FRB
are narrowly collimated means that the failure to observe a FRB coincident
with a gravitational wave event would not disprove the hypothesis.
\section{Data Availability}
This theoretical study did not generate any new data.

\label{lastpage} 

\begin{thebibliography}{99}
	\bibitem[\protect\citeauthoryear{Bera \& Chengalur}{2019}]{BC19}
		Bera, A. \& Chengalur, J. N. 2019 \mnras\ 490, L12.
	\bibitem[\protect\citeauthoryear{Bobrick {\it et al.\/}}{2017}]{B17}
		Bobrick, A., Davies, M. B. \& Church, R. P. 2017 \mnras\
		467, 3556.
	\bibitem[\protect\citeauthoryear{Bobrick {\it et al.\/}}{2021}]{B21}
		Bobrick, A., Zenati, Y., Perets, H. B. {\it et al.\/} 2021
		arXiv:2104.03415.
	\bibitem[\protect\citeauthoryear{Bochenek {\it et al.\/}}{2020a}]
		{B20a} Bochenek, C. D., McKenna, D. L., Belov, K. V. {\it et
		al.\/} 2020a \pasp\ 132, 034202.
        \bibitem[\protect\citeauthoryear{Bochenek {\it et al.\/}}{2020b}]
                {B20b} Bochenek, C., Kulkarni, S. Ravi, V. McKenna, D.,
                Hallinan, G. \& Belov, K. 2020b arXiv:2005.10828.
	\bibitem[\protect\citeauthoryear{Chen {\it et al.\/}}{2021}]{CTHC21}
		Chen, H.-L., Tauris, T. M., Han, Z. {\it et al.\/} 2021
		\mnras\ 503, 3540.
	\bibitem[\protect\citeauthoryear{CHIME/FRB Collaboration}{2020a}]
                {CHIME20a} CHIME/FRB Collaboration 2020a \nat\ 582, 351
                (arXiv:2001.10275).
	\bibitem[\protect\citeauthoryear{CHIME/FRB Collaboration}{2020b}]
                {C20b} CHIME/FRB Collaboration 2020b arXiv:2005.10324.
	\bibitem[\protect\citeauthoryear{Clark}{1975}]{C75} Clark, G. W.
		1975 \apjl\ 199, L143.
	\bibitem[\protect\citeauthoryear{Conselice {\it et al.\/}}{2016}]
		{C16} Conselice, C. J., Wilkinson, A., Duncan, K. {\it et
		al.\/} 2016 \apj\ 830, 83.
        \bibitem[\protect\citeauthoryear{Cruces {\it et al.\/}}{2021}]{C21}
                Cruces, M., Spitler, L. G., Scholz, P. {\it et al.\/} 2021
                \mnras\ 500, 448 (arXiv:2008.03461).
	\bibitem[\protect\citeauthoryear{Dokuchaev \& Eroshenko}{2017}]
		{DE17} Dokuchaev, V. I. \& Eroshenko, Yu. N. 2017
		arXiv:1701.02492.
	\bibitem[\protect\citeauthoryear{Ferrario, de Martino \&
		G{\"a}nsicke}{2015}]{FMG15} Ferrario, L., de Martino, D. \&
		G{\"a}nsicke, B. T. 2015 Sp.~Sci.~Rev. 191, 111.
	\bibitem[\protect\citeauthoryear{Gajjar {\it et al.\/}}{2018}]{G18}
		Gajjar, V., Siemion, A. V. P., Price, D. C. {\it et al.\/}
		2018 \apj\ 863, 2.
	\bibitem[\protect\citeauthoryear{Joss, Katz \& Rappaport}{1979}]
		{JKR79} Joss, P. C., Katz, J. I. \& Rappaport, S. A. (1979)
		\apj\ 230, 176.
	\bibitem[\protect\citeauthoryear{Katz}{1975}]{K75} Katz, J. I. 1975
		Nature 253, 698.
	\bibitem[\protect\citeauthoryear{Katz}{1982}]{K82} Katz, J. I. 1982
		\apj\ 260, 371.
	\bibitem[\protect\citeauthoryear{Katz}{1996}]{K96} Katz, J. I. 1996
		\apj\ 463, 305.
	\bibitem[\protect\citeauthoryear{Katz}{2017}]{K17} Katz, J. I. 2017
		\mnras\ 467, L96.
	\bibitem[\protect\citeauthoryear{Katz}{2018}]{K18} Katz, J. I. 2018
		Prog. Part. Nucl. Phys. 103, 1.
	\bibitem[\protect\citeauthoryear{Katz}{2019}]{K19} Katz, J. I. 2019
		\mnras\ 487, 491.
	\bibitem[\protect\citeauthoryear{Katz}{2020a}]{K20a} Katz, J. I.
		2020a \mnras\ 494, L64.
	\bibitem[\protect\citeauthoryear{Katz}{2020b}]{K20b} Katz, J. I.
		2020b \mnras\ 499, 2319.
	\bibitem[\protect\citeauthoryear{Katz}{2020c}]{K20c} Katz, J. I.
		2020c \mnras\ 494, 3463.
	\bibitem[\protect\citeauthoryear{Katz}{2021a}]{K21a} Katz, J. I.
		2021a \mnras\ 501, L76.
	\bibitem[\protect\citeauthoryear{Katz}{2021b}]{K21b} Katz, J. I.
		2021b arXiv:2106.05212.
	\bibitem[\protect\citeauthoryear{Kirsten {\it et al.\/}}{2021}]
		{Ki21} Kirsten, F., Marcote, B., Nimmo, K. {\it et al.\/}
		2021 arXiv:2105.11445.
	\bibitem[\protect\citeauthoryear{Kremer, Piro \& Li}{2021}]{KPL21}
		Kremer, K., Piro, A. L. \& Li, D. 2021 \apjl\ in press
		arXiv:2107.03394.
        \bibitem[\protect\citeauthoryear{Li {\it et al.\/}}{2020}]{Li20}
                Li, C. K., Lin, L., Xiong, S.-L. {\it et al.\/} 2020
                arXiv:2005.11071.
	\bibitem[\protect\citeauthoryear{Li {\it et al.\/}}{2021}]{Li21}
		Li, D., Wang, P., Zhu, W. W. {\it et al.\/} 2021
		arXiv:2107.08205.
	\bibitem[\protect\citeauthoryear{Lu, Beniamini \& Kumar}{2021}]
		{LBK21} Lu, W., Beniamini, P. \& Kumar, P. 2021 \mnras\
		submitted arXiv:2107.04059.
	\bibitem[\protect\citeauthoryear{Margalit, Metzger \& Sironi}{2020}]
		{Ma20} Margalit, B., Metzger, B. D. \& Sironi, L. 2020
		\mnras\ 494, 4627.
        \bibitem[\protect\citeauthoryear{Mereghetti {\it et al.\/}}{2020}]
                {M20} Mereghetti, S., Savchenko, V., Gotz, D. {\it et al.\/}
                2020 \apjl\ 898, L29 arXiv:2005.06335.
	\bibitem[\protect\citeauthoryear{Mottez \& Zarka}{2014}]{MZ14}
		Mottez, F. \& Zarka, P. 2014 \aap\ 569, 86.
	\bibitem[\protect\citeauthoryear{Mottez, Zarka \& Voisin}{2020}]
		{MZV20} Mottez, F., Zarka, P. \& Voisin, G. 2020 \aap\ 644,
		145.
        \bibitem[\protect\citeauthoryear{Pastor-Marazuela {\it et al.\/}}
                {2020}]{PM20} Pastor-Marazuela, I., Connor, L., van Leeuwen,
                J. {\it et al.\/} 2020 arXiv:2012.08348.
        \bibitem[\protect\citeauthoryear{Pilia {\it et al.\/}}{2020}]{Pi20}
                Pilia, M., Burgay, M., Possenti, A. {\it et al.\/} 2020
                \apjl\ 896, L40 (arXiv:2003.12748).
	\bibitem[\protect\citeauthoryear{Platts {\it et al.\/}}{2019}]{P19}
		Platts, E., Wellman, A., Walters, A. {\it et al.\/} 2019
		Phys. Rep. 821, 1 \url{frbtheorycat.org} accessed June 6,
		2021.
        \bibitem[\protect\citeauthoryear{Pleunis {\it et al.\/}}{2020}]
                {Pl20} Pleunis,~Z., Michilli,~D., Bassa,~C.~G. {\it et
                al.\/} 2020 arXiv:2012.08372.
	\bibitem[\protect\citeauthoryear{Pooley {\it et al.\/}}{2003}]{P03}
		Pooley, D., Lewin, W. H. G., Anderson, S. F. {\it et al.\/}
		2003 \apjl\ 591, L131.
        \bibitem[\protect\citeauthoryear{Rajwade {\it et al.\/}}{2020}]{R20}
                Rajwade, K. M., Mickaliger, M. B., Stappers, B. W. {\it et
                al.\/} 2020 \mnras\ 495, 3551 (arXiv:2003.03596).
        \bibitem[\protect\citeauthoryear{Ridnaia {\it et al.\/}}{2020}]{Ri20}
                Ridnaia, A., Svinkin, D., Fredericks, D. {\it et al.\/} 2020
                arXiv:2005.11178.
	\bibitem[\protect\citeauthoryear{Srinivasan}{2010}]{S10}
		Srinivasan, G. 2010 New Astron. Rev. 54, 93.
        \bibitem[\protect\citeauthoryear{Tavani {\it et al.\/}}{2020}]{T20}
                Tavani, M., Casentini, C., Ursi, A. {\it et al.\/} 2020
                arXiv:2005.12164.
	\bibitem[\protect\citeauthoryear{Totani}{2013}]{T13} Totani, T. 2013
		\pasj\ 65, L12.
	\bibitem[\protect\citeauthoryear{van Haaften {\it et al.\/}}{2011}]
		{vH11} van Haaften, L. M., Nelemans, G., Voss, R. {\it et
		al.\/} 2011 \aap\ 537A, 104.
	\bibitem[\protect\citeauthoryear{Vigan\`{o} {\it et al.\/}}{2013}]
		{V13} Vigan\`{o}, D., Rea, N., Pons, J. A. {\it et al.\/}
		2013 \mnras 434, 123.
	\bibitem[\protect\citeauthoryear{Wang {\it et al.\/}}{2016}]{W16}
		Wang, J.-S., Yang, Y.-P., Wu, X.-F. {\it et al.\/} 2016
		\apj\ 822, L7.
	\bibitem[\protect\citeauthoryear{Yu {\it et al.\/}}{2021}]{Y21} Yu,
		S., Lu, Y. \& Jeffery, C. S. 2021 arXiv:2103.01884.
	\bibitem[\protect\citeauthoryear{Zenati {\it et al.\/}}{2020}]{Z20}
		Zenati, Y., Bobrick, A., Perets, H. B. 2020 \mnras\ 493,
		3956.
	\bibitem[\protect\citeauthoryear{Zhang {\it et al.\/}}{2018}]{Z18}
		Zhang, Y.-G., Gajjar, V., Foster, G. {\it et al.\/} 2018
		\apj\ 866, 149.
\end{thebibliography}
\end{document}